\documentclass{article}
\usepackage[utf8]{inputenc}

\title{COVID_BOA}
\author{e.o.pyzerknapp }
\date{May 2020}

\usepackage{graphicx}

\usepackage{float}
\usepackage{booktabs}
\usepackage{tabularx}
\usepackage{authblk}
\begin{document}

\title{Using Bayesian Optimization to Accelerate Virtual Screening for the
Discovery of Therapeutics Appropriate for Repurposing for COVID-19}

\author[1]{Ed Pyzer-Knapp}%
\affil[1]{IBM Research Europe}%

\vspace{-1em}

  \date{\today}

\begingroup
\let\center\flushleft
\let\endcenter\endflushleft
\maketitle
\endgroup

\begin{abstract}
The novel Wuhan coronavirus known as SARS-CoV-2 has brought almost
unprecedented effects for a non-wartime setting, hitting social,
economic and health systems hard.~ Being able to bring to bear
pharmaceutical interventions to counteract its effects will represent a
major turning point in the fight to turn the tides in this ongoing
battle.~ Recently, the World's most powerful supercomputer, SUMMIT, was
used to identify existing small molecule pharmaceuticals which may have
the desired activity against SARS-CoV-2 through a high throughput
virtual screening approach.~ In this communication, we demonstrate how
the use of Bayesian optimization can provide a valuable service for the
prioritisation of these calculations, leading to the accelerated
identification of high-performing candidates, and thus expanding the
scope of the utility of HPC systems for time critical screening.%
\end{abstract}%

\section*{Introduction}

{\label{874460}}

The virus now known as SARS-CoV-2 seems to have initiated in Wuhan, in
Hubei Province, China at the end of 2019.\cite{li2020new, xu2020evolution} It has since
become a global pandemic, impacting the economic, social and health
systems of almost every country in the world.~ In a race to discover
potential pharmaceutical interventions, some researchers have turned to
high-throughput virtual screening as an avenue to prioritise the
screening of existing commercially available or pharma-internal
proprietary, compounds for activity against the virus.~ Screening at
scale has been enabled by an unprecedented communal activity known as
the COVID-19 HPC consortium \cite{xsede}, which includes systems
provided by

\begin{itemize}
\item
  U.S. Department of Energy (DOE) Advanced Scientific Computing Research
  (ASCR)
\item
  U.S. DOE National Nuclear Security Administration (NNSA)
\item
  Rensselaer Polytechnic Institute
\item
  MIT/Massachusetts Green HPC Center (MGHPCC)
\item
  IBM Research WSC
\item
  U.S. National Science Foundation (NSF)
\item
  NASA High-End Computing Capability
\item
  Amazon Web Services
\item
  Microsoft Azure Cloud and High Performance Computing
\item
  Hewlett Packard Enterprise, and
\item
  Google
\end{itemize}

It is anticipated that there will be high demand for the use of HPC
resources such as are provided by the consortium, and so an ability to
`power up' smaller compute clusters to accelerate the (re-)discovery of
pharmaceutical interventions through the use of in silico methods could
prove very valuable.~ In this paper we propose the use of Bayesian
optimization as such a technology, and demonstrate its potential through
a demonstration on an early dataset collected on the IBM SUMMIT
supercomputer by Smith, et al.\cite{Smith_2020}

\section*{High Throughput Virtual
Screening}

{\label{303353}}

High throughput virtual screening (HTVS) has emerged as a powerful
methodology for attacking the identification of candidate molecules for
materials and pharmaceutical applications.\cite{Pyzer_Knapp_2015, Langer_2001, Good_2000}. Benefit
over traditional high throughput methods, such as robotic
assays, mainly stem from two key areas:

\begin{enumerate}
\item
  Once there is a working `virtual protocol' for an experiment, scaling
  that experiment is mainly driven through access to computational
  resources.
\item
  A virtual high-throughput method is not limited to considering
  molecules which exist in the lab, and thus does not differentiate
  molecules based on synthetic complexity or stability.
\end{enumerate}

The cloud, in particular, has driven more and more individuals and
organisations to have access to sufficient computational resource to
perform HTVS techniques, at least at reasonable
timescale.\cite{Ol_a__2019, Ellingson_2012, Dolezal_2015, Capuccini_2017} It should be worth noting, however, that
in many cases, the performance of the cloud is still insufficient for some
HPC applications.\cite{Netto_2018} For some problems, however, the
urgency of the outcome dictates that a scale of computing and
performance currently not yet feasible in the cloud be used.~ One such
example of this is the work of Jeremy Smith et al, who used the world's
fastest supercomputer, Summit \cite{sites} to screen over 8,000
known biologically active compounds, including many approved drug
compounds, for potential activity against the SARS-CoV-2 virus~through a
study of their binding of the ACE-2 receptor.\cite{Smith_2020} ~ Smith
used restrained temperature molecular dynamics to locate six different
configurations of the host protein and then used the molecular docking
code autodock-VINA~\cite{Trott_2009} to score potential poses of the
small molecule pharmaceuticals included in the SWEETLEAD
library.\cite{Novick_2013} The use of the SUMMIT supercomputer was
deemed necessary to reduce the time to score the entire library of more
than 8,000 compounds in days, rather than the more normally expected
months.~

\section*{Bayesian Optimization}

{\label{507127}}

Bayesian optimization is a black-box optimization algorithm commonly
used in cases where the evaluation of data-points is computationally or
financially expensive, or in other situations in which the number of
evaluations of potential configurations is required to be minimized.~
It has become popular recently in the machine learning community for the
tuning of hyper-parameters of machine learning models, but has also
demonstrated significant ability to accelerate scientific tasks such as
the discovery of new materials~\cite{spac2017}, pharmaceutically
active ingredients \cite{Pyzer_Knapp_2018} and the parameterisation of
classical force-fields. \cite{McDonagh_2019}

\begin{figure}[H]
\begin{center}
\includegraphics[width=0.70\columnwidth]{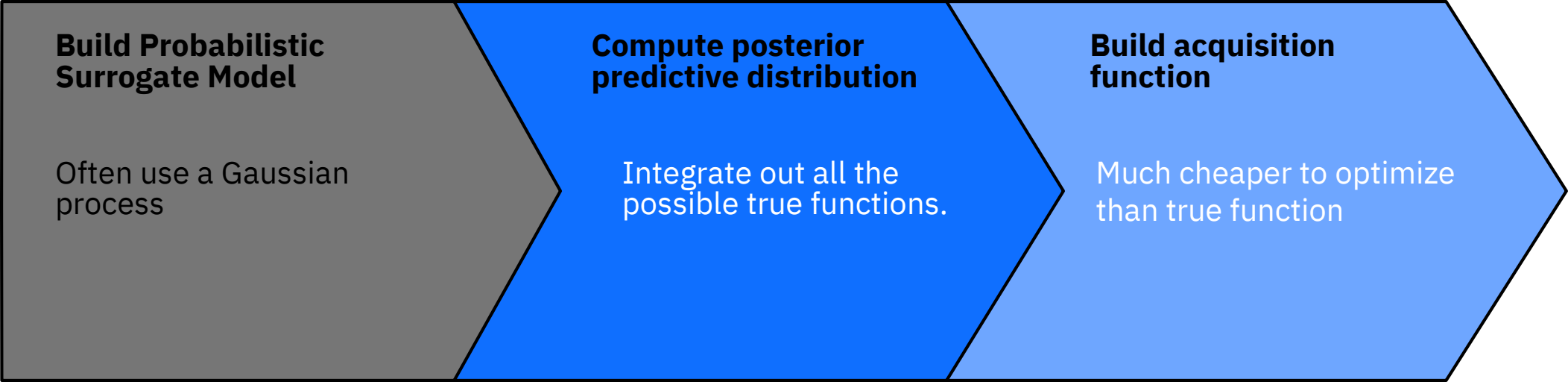}
\caption{{The stages of the Bayesian optimization algorithm
{\label{692099}}%
}}
\end{center}
\end{figure}

At a high level, the algorithm follows a relatively simple flow, which
is shown in Figure~{\ref{692099}}. The lack of access
to analytical forms is mitigated through the use of a probabilistic
surrogate model, often a Gaussian process \cite{snoek2012practical, brochu2010tutorial} although
other Bayesian models can be used.\cite{spac2017}

Once the model has been used to calculate predictive means and variances
(uncertainties) it is possible to calculate a value commonly termed
the~\emph{improvement:}

\emph{\[\gamma= \frac{y_{pred} - f^{*}}{\sigma^{2}}\]}

where~\(y_{pred}\) represents a prediction from the Bayesian
model,~\(\sigma^{2}\) represents the corresponding model
uncertainty, and~\(f^{*}\) represents the best value found thus
far.~ This value can then be treated with a number of strategies, known
as acquisition functions, which determine how valuable each data point
is to acquire.~ Perhaps the most commonly used acquisition function is
known as Expected Improvement \cite{Mo_kus_1975} and takes the
following form:

\[EI(x) = \mu(x) - f^{*}\Phi(\gamma) + \sigma(x) \phi(\gamma)\]

where~\(\Phi(\gamma)\) and~\(\phi(\gamma)\) represent the CDF and
PDF of the standard normal distribution applied to the improvement
function.~~

A challenge with the Expected Improvement methodology is that it is an
inherently serial algorithm, since in order to generate a new posterior,
there must be new information added.~ Several methodologies have evolved
to tackle this problem, whether through partitioning the acquisition
function, for example using K-means \cite{groves2018efficient} or through
penalising the acquisition function from selecting certain
points. \cite{gonzalez2016batch}. In this study we use a batch variant of
Thompson sampling, PDTS, \cite{spac2017} to achieve the required
parallel acquisition necessary to exploit high performance computing
clusters.~ PDTS was chosen in part for its excellent performance in the
task of performing HTVS on~ large materials databases, and in part for
its excellent scalability to input spaces with large numbers of
dimensions, which are frequently found in chemical problems.~

\section*{Methodology}

{\label{662362}}

For this proof of concept study, simulations were not run on the SUMMIT
supercomputer, but instead the openly released results of Jeremy Smith,
et al. \cite{Smith_2020} were used as an oracle. There is, of course, no
reason why this could not be repeated on new problems using a resource
such as SUMMIT, especially as Smith et al. have open-sourced the
deployment scripts necessary.~ For the purposes of this study, we
targeted the two situations described in \cite{Smith_2020} as two
minimization challenges, with the aim being to locate particularly low
VINA scores efficiently.~ We understand that VINA scores are not a
perfect indicator of performance, and thus entreat the reader to
consider the optimization as a prioritization exercise - there is
nothing to stop the entire library being calculated, but clearly the
order in which these calculations happen can have a significant effect
on the speed of discovery of candidates for re-purposing.

Each molecule was described by an extended connectivity fingerprint
(ECFP), with a radius of 2, and a 512 bit hashing, as this was
sufficient to ensure that there were no clashes between molecules in the
library.\cite{rogers2010extended} Clashes arise when two molecules are given
the same fingerprint, and so choosing a fingerprint length is a trade-off
between fingerprint length (long fingerprints are less likely to clash)
and computational cost (long fingerprints make the machine learning more
expensive, and harder).

A Gaussian process model was used for the construction of the
acquisition function, using a Matern kernel of the form:

\[K_{matern} = \sigma ^{2}\left(1+{\frac {{\sqrt {5}}d}{\rho }}+{\frac {5d^{2}}{3\rho ^{2}}}\right)\exp \left(-{\frac {{\sqrt {5}}d}{\rho }}\right)\]

where the hyperparameters of the kernel are optimized to maximise the
log marginal likelihood of the model.

The PDTS algorithm was used to generate batches of molecules to send to
the oracle. For this application,~ a batch size of 10 was used to
balance the granularity of updating the posterior distribution regularly
with the obvious efficiency benefit to operating in parallel.~ The PDTS
algorithm was seeded with 15 randomly selected molecules, and then run
for 50 epochs, resulting in a set of just over 500 molecules.~ This set
size was chosen as it was estimated that more commonly available compute
capability could run a set of this size in the same number of days that
SUMMIT took to run the full set of 8,000 molecules.

\par\null

\section*{Results and Discussion}

{\label{996869}}

In order to evaluate the effectiveness of the Bayesian optimization
approach, we believe there are two key axes to evaluate:

\begin{enumerate}
\item
  How rapidly does the approach discover highly active molecules (as
  represented by highly negative VINA scores)
\item
  How enriched are the samples taken by the Bayesian optimization
  approach, as compared to the exhaustive set (i.e. how many more
  samples are in the `highly active' bins than might be expected given
  the known distribution of VINA scores)
\end{enumerate}

We attack this problem for both situations outlined by Smith et al,
namely docking into the interface and isolated viral S-protein.~ After
the budget of 50 epochs has been exhausted, we interrogate the set which
would have been calculated, and compare it to randomly sampling from the
set for the same amount of time.~ This experiment is repeated 5 times to
generate uncertainty estimates resulting from different seeding of the
sampler, and to demonstrate the volatility of the random sampling
approach.~

\subsection*{Optimization of Vina
Score}

{\label{584580}}

For this first task, the simple challenge of locating molecules with a
desirable VINA score is considered.~ It is understood that this is not
the only consideration for which a successful active pharmaceutical may
be identified for further testing for efficacy against SARS-CoV-2, but
rapidly identifying candidates with this property is a reasonable way of
prioritising the spend of compute resource.~ Additionally, if it can be
shown that the Bayesian optimization methodology is able to optimize to
this task, it is reasonable to assume that there is likely to be
advantage to using the same methodology for similar derived tasks, which
may more closely map to the users actual needs.

\begin{figure}[H]
\begin{center}
\includegraphics[width=\columnwidth]{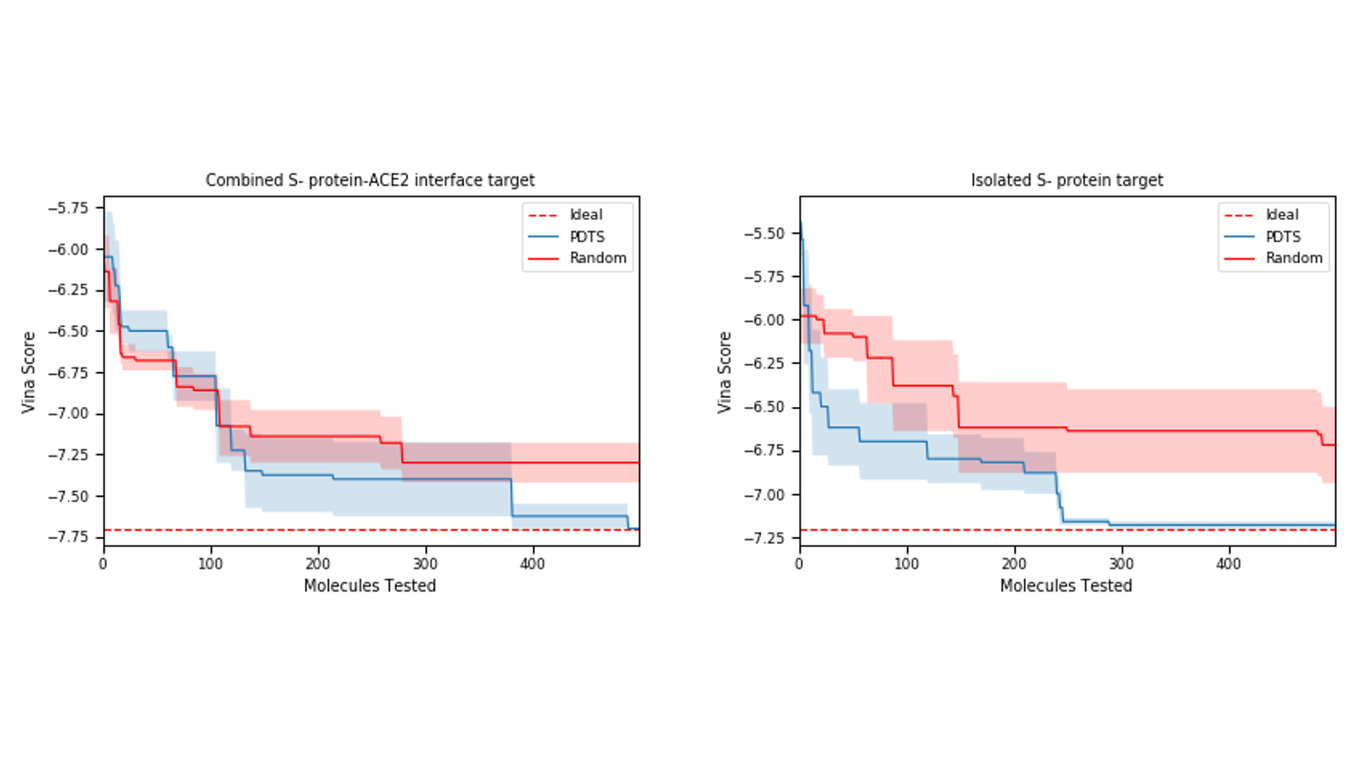}
\caption{{Optimization progress as a function of molecules tested for the combined
S-protein-ACE2 interface target (left) and the Isolated S-protein target
(right).~ Confidence intervals were calculated from five repeats using
the bootstrap methodology.
{\label{235636}}%
}}
\end{center}
\end{figure}

It can be seen from Figure~{\ref{235636}} that the PDTS
sampler - the selected Bayesian optimization methodology for this study
- significantly outperformed the random search.~ It achieved this both
in terms of the targets obtained - locating the global optimum in
\textless{}500 steps in almost all of the 10 runs (2 sets of 5
replicates), and doing so reliably, with significantly lower
uncertainties.~ This is strong evidence that Bayesian optimization meets
the first criteria for an effective prioritisation tool for HTVS
activities.

\subsection*{Enrichment of Sampling
Sets}

{\label{304690}}

It is also important to consider not just the single optimium point, but
a set of molecules with desirable properties, which affords the user
some options for further investigation.~ There are a number of ways to
consider the enrichment, but here we will consider two metrics:

\begin{enumerate}
\item
  How the average score of the top 10 ranked molecules sampled thus far
  evolves as the sampling progresses.
\item
  Where the `sampling density' is concentrated after sampling 500
  molecules, as compared to the overall distribution of molecules
\end{enumerate}

If the Bayesian optimization methodology shows significantly more
desirable behaviour than the random search method then this is further
proof that this methodology can be utilised to prioritise HVTS screens
successfully.

\begin{figure}[H]
\begin{center}
\includegraphics[width=\columnwidth]{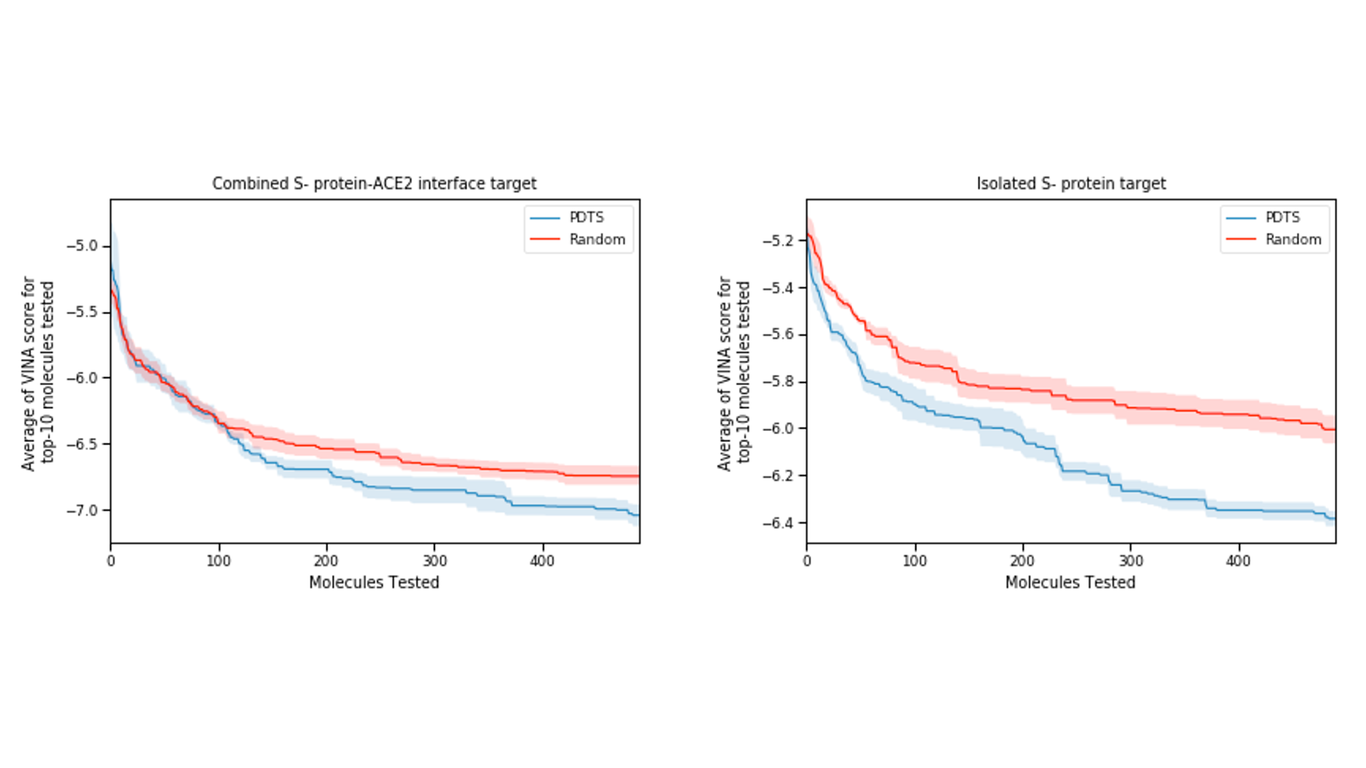}
\caption{{Evolution of the average VINA score of the top 10 ranked molecule for
the PDTS sampler (blue) and a random sample (red).~ These were generated
for the combined S-protein-ACE2 interface target (left) and the Isolated
S-protein target (right).~ Confidence intervals were calculated from
five repeats using the bootstrap methodology.
{\label{787847}}%
}}
\end{center}
\end{figure}

Figure~{\ref{787847}} shows the evolution of the
average VINA score of the top 10 candidates over the 50-epoch
optimization (500 molecules sampled).~ As can be seen, in both tasks,
the PDTS sampler robustly achieves a lower top10 score than the random
sampler, indicating that higher quality molecules are being sampled.

\begin{figure}[H]
\begin{center}
\includegraphics[width=\columnwidth]{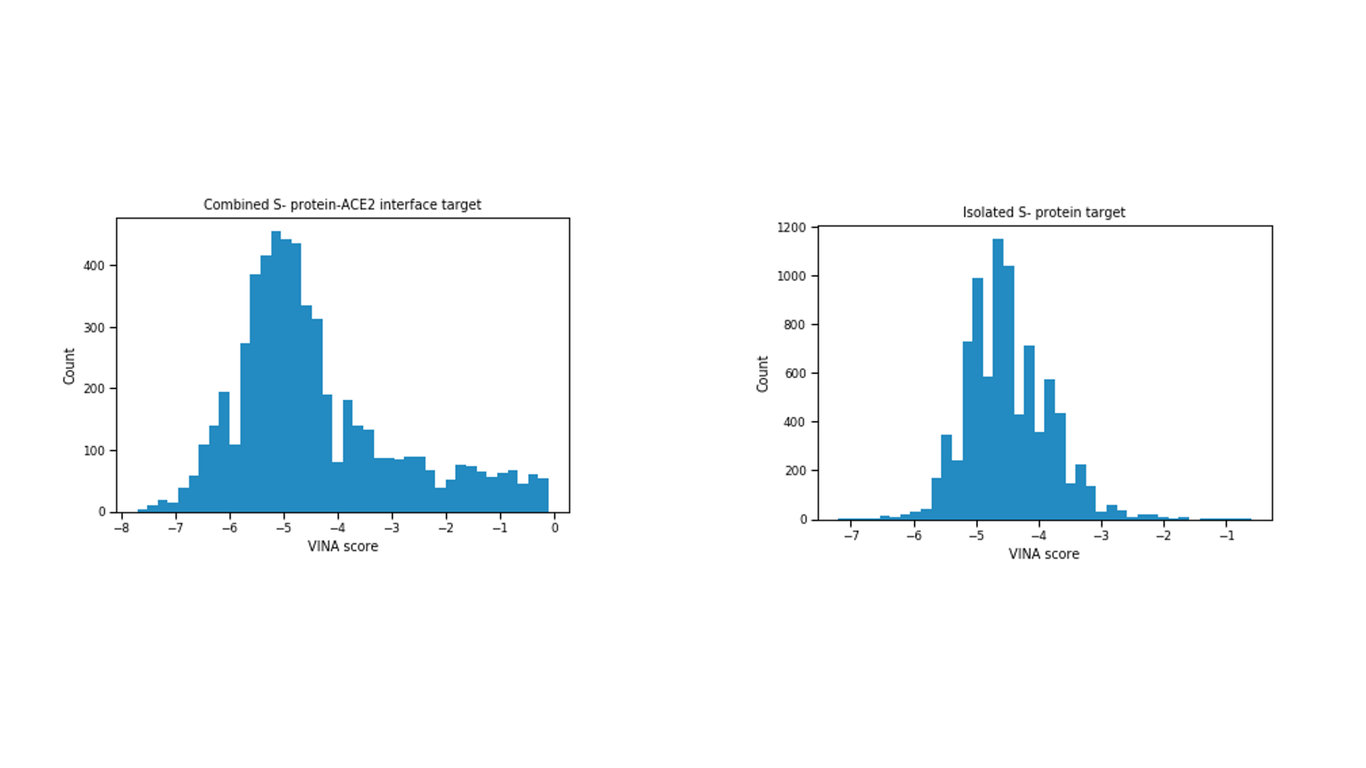}
\caption{{Histogram of the VINA scores calculated by Smith et al for the two tasks
considered.
{\label{497727}}%
}}
\end{center}
\end{figure}

It can be seen by looking at the distribution of VINA scores for the
whole set of molecules considered by Smith et
al,(Figure~{\ref{497727}}) that there are simply not
many molecules in the set which display a desirable VINA score.~
Additionally drop off in the number of molecules displaying a desirable
VINA scores is more marked for the Isolated S-protein target than for
the Combined S-protein-ACE2 interface target.~ Intuititvely, therefore,
it is unreasonable to expect a randomly drawn subset of these molecules
to display strong potential as candidates against SARS-CoV-2.~ By
comparing the distribution of a subset of the overall sets which
correspond to desirable VINA scores against the distrubtion of molecules
selected by the Bayesian PDTS sampler, we can determine the `enrichment'
the Bayesian sampler offers.

\begin{figure}[H]
\begin{center}
\includegraphics[width=\columnwidth]{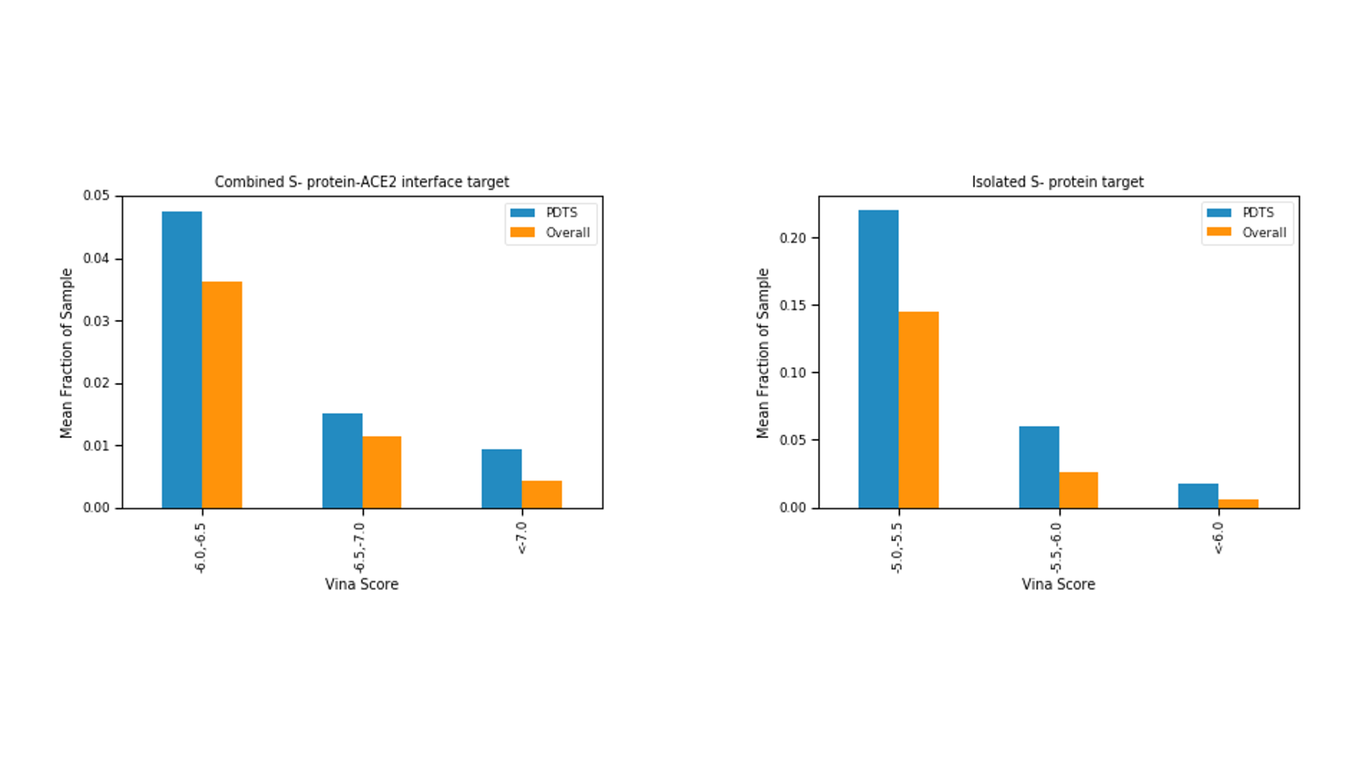}
\caption{{Comparison of the distribution of VINA scores in the sampled set by PDTS
and the overall distribution displayed by the full set as collected by
Smith, et al. For each task, the bins were determined to capture the top
20\% of the overall distribution.~ PDTS distributions were calculated as
an average over the 5 replicate experiments.~
{\label{183592}}%
}}
\end{center}
\end{figure}

Figure~{\ref{183592}} shows this data averaged over the
five replicate runs of PDTS. . Here, we plot the `mean fraction of
sample' as a scoring metric.~ This is simply the mean value (over the
five runs for PDTS) of the fraction of the sample which would be placed
within this bin.~ This is compared to the fraction which is observed
within these bins for the full set of molecules.~ The fact that in all
cases, the blue bar (representing the PDTS approach) is greater than the
orange bar (representing the overall distribution) demonstrates the
effectiveness of the PDTS methodology to draw enriched sets of molecules
for testing, and is further proof of the efficiency of the Bayesian
optimization methodology to prioritise tasks for HVTS.~

Table~{\ref{245009}} indicates numerically the
performance of the PDTS algorithm for these three tasks. It can be seen
than on every task the PDTS method of Bayesian optimization outperforms
random sampling and thus shows great potential for the efficient
prioritisation of HTVS testing of pharmaceuticals for activity against
SARS-CoV-2.~ This is both demonstrated by the~ ability of the PDTS
sampler to efficiently locate the global optima in the two tasks, but
also to do so reliably - as indicated by the low standard deviations
over runs.~ We also demonstrate the ability of the PDTS sampler to
create enriched samples on limited budgets, which was demonstrated by
the greater scores in both the Top10 average test, and the Top `Bin'
percentage test.~ Again, both of these tests display reasonable standard
deviations across runs, demonstrating the reproducibility of the
methodology.
\begin{table}[H]
\centering
\begin{tabularx}{\columnwidth}{XXXXX|X}
\toprule
\multicolumn{6}{c}{\textbf{Methodology}}\\

  & PDTS & (STD) & Random & (STD) & Ideal \\
  \toprule
& \multicolumn{4}{c}{\textit{Combined S- protein-ACE2 interface target}}&\\
Best VINA Score Sampled  & \textbf{-7.7} & 0.14 & -6.74 & 0.28 & -7.7  \\
Top 10 Average  & \textbf{-7.04} & 0 & -7.3 & 0.16 & -7.54  \\
Percentage in top 'bin'  & \textbf{0.92} & 0.42 & 0.42 & NA & NA \\
\bottomrule
% & \multicolumn{5}{>{\hsize=\dimexpr 5\hsize+5\tabcolsep+\arrayrulewidth}X|}{Isolated S-protein target} \\
& \multicolumn{4}{c}{\textit{Isolated S-protein target}}* \\
Best VINA Score Sampled   & -\textbf{7.18} & 0.04 & -6.72 & 0.47 & -7.18 \\
Top 10 Average  & \textbf{-6.39} & 0.07 & -6.01 & 0.12 & -6.81 \\
Percentage in top 'bin'&  \textbf{1.79} & 0.33 & 0.56 & NA & NA \\
\bottomrule
\end{tabularx}
\caption{{Overall results of the three metrics for investigation as to the
potential for Bayesian optimization to act as an efficient prioritisation
engine for HTVS.~ For `Percentage in top bin', the `Random' value refers
to the overall percentage, rather then a particular random sample.~ For
`Best VINA' and `Top10' metrics,~ PDTS and random results are averaged
over 5 runs, with the standard deviation of those runs also being
reported.
{\label{245009}}%
}}
\end{table}\section*{Conclusion}

{\label{611140}}

When concluding this study, it is first important to emphasise what this
paper is not.~ It is not an exhaustive benchmark of optimization
methodologies for application to HTVS.~ Indeed, the no-free-lunch
theorem says that there is not one globally guaranteed optimal
methodology existing, anyway. We have shown in the
past\cite{Pyzer_Knapp_2018} that Bayesian optimization offers significant
benefits over a greedy methodology, which is prone to catastrophic local
optimization, and so we do not seek to reproduce those results here.
What this study demonstrates is that, for problems which are both time
critical and compute intensive, Bayesian optimization (specifically here
the PDTS algorithm) have significant benefits when applied to HTVS
screens.~ With a global problem such as the SARS-CoV-2 pandemic, speed
to solution is necessary, but must also be tempered with a pragmatic
approach to resource allocation.~ There are simply not enough SUMMIT
supercomputers in the world for everyone who might want to use them.~
Here, we demonstrate that for the challenge attempted by Smith, et al.
for determining a set of molecules for further investigation using the
HTVS methodology, promising results could be reached with a smaller
compute cluster, if paired with a prioritisation mechanism powered by
Bayesian optimization.

\par\null

\section*{Acknowledgements}

The author would like to thank Jeremy Smith for making his data
publicly available, and to Dave Turek, Kirk Jordan and Chris Porter of
IBM Corp, for their helpful discussions, and insight.
Additionally, the author would like to take this opportunity to thank those who are working on the front line to respond to the SARS-CoV-2 global pandemic. 

\section*{References}
\bibliographystyle{acm}
\bibliography{references}
\end{document}